\documentstyle[10pt,epsf,epsfig,dp_delphititle,amssymb]{dp_delphi}
\makeindex
\def\DpPaperGroup{EP}
\def\DpPaperRef{2000-007 \\ Revised Version}
\def\DpDate{7 June 2000}
\def\DpAuthors{DELPHI Collaboration}
\def\DpSubmit{(Eur. Phys. J. C17(2000)207)}
\def\DpTitle{{Identified Charged Particles  
              in Quark and Gluon Jets}}
\def\DpComment{ }
\def\DpEMail{ }

\renewcommand{\rb}[1]{\raisebox{1.5ex}[-1.5ex]{#1}}
\newcommand{\rul}{\rule[-3mm]{0mm}{8mm}}

\newcommand{\mc}{\multicolumn}
\newcommand{\m}{\mbox}
\newcommand{\fileid}[1]{#1.eps} 
\newcommand{\tref}[1]{\mbox{{Table~}\ref{#1}}}
\newcommand{\fref}[1]{\mbox{{Figure~}\ref{#1}}}

\newlength{\bwid}
\newlength{\fwid}

\newlength{\breite}
\newlength{\breitemp}
\newcommand{\epem} {\ifmmode{e^+e^-}\else{$e^+e^-$}\fi}
\newcommand{\qqbar} {\ifmmode{q\bar{q}}\else{$q\bar{q}$}\fi}
\def\z{${ Z}$\ }
\def\p{$p$\ }
\def\pie{$\pi^+$\ }
\def\kp{$K^+$\ }
\def\b{$b$\ }
\def\bbg{$b\bar{b}g$\ }
\newcommand{\eps}{{\ifmmode \varepsilon \else $\varepsilon$\fi}}
\newcommand{\thetac}{{\ifmmode \theta_C\else $\theta_C$\fi}}

\def\gevc{${GeV}/c$\ }

\newcommand{\qgj}{quark and gluon jets }
\newcommand{\gqj}{gluon and quark jets }
\newcommand{\JS}{{\sc Jetset}}
\newcommand{\AR}{{\sc Ariadne}}
\newcommand{\HW}{{\sc Herwig}}
\newcommand{\BC}{\begin{center}}
\newcommand{\EC}{\end{center}}
\newcommand{\BE}{\begin{equation}}
\newcommand{\EE}{\end{equation}}
\newcommand{\BEA}{\begin{eqnarray}}
\newcommand{\EEA}{\end{eqnarray}}
\newcommand{\BA}{\begin{array}}
\newcommand{\EA}{\end{array}}
\newcommand{\BI}{\begin{itemize}}
\newcommand{\EI}{\end{itemize}}
\newcommand{\BF}{\begin{figure}}
\newcommand{\EF}{\end{figure}}
\newcommand{\BT}{\begin{table}}
\newcommand{\ET}{\end{table}}
\newcommand{\BTB}{\begin{tabular}}
\newcommand{\ETB}{\end{tabular}}
\newcommand\BM{\begin{minipage}}
\newcommand\EM{\end{minipage}}

\newcommand{\Erg}[3]{\ifmmode{#1\pm#2_{stat.}\pm#3_{sys.}}\else{$#1\pm#2_{stat.}\pm#3_{sys.}$}\fi}
\newcommand{\erg}[3]{\ifmmode{\scriptstyle#1\,\pm\,#2\,\pm\,#3}\else{$\scriptstyle#1\,\pm\,#2\,\pm\,#3$}\fi}
\begin{document}
\makeatletter
\newcount\@tempcntc
\def\@citex[#1]#2{\if@filesw\immediate\write\@auxout{\string\citation{#2}}\fi
  \@tempcnta\z@\@tempcntb\m@ne\def\@citea{}\@cite{\@for\@citeb:=#2\do
    {\@ifundefined
       {b@\@citeb}{\@citeo\@tempcntb\m@ne\@citea\def\@citea{,}{\bf ?}\@warning
       {Citation `\@citeb' on page \thepage \space undefined}}%
    {\setbox\z@\hbox{\global\@tempcntc0\csname b@\@citeb\endcsname\relax}%
     \ifnum\@tempcntc=\z@ \@citeo\@tempcntb\m@ne
       \@citea\def\@citea{,}\hbox{\csname b@\@citeb\endcsname}%
     \else
      \advance\@tempcntb\@ne
      \ifnum\@tempcntb=\@tempcntc
      \else\advance\@tempcntb\m@ne\@citeo
      \@tempcnta\@tempcntc\@tempcntb\@tempcntc\fi\fi}}\@citeo}{#1}}
\def\@citeo{\ifnum\@tempcnta>\@tempcntb\else\@citea\def\@citea{,}%
  \ifnum\@tempcnta=\@tempcntb\the\@tempcnta\else
   {\advance\@tempcnta\@ne\ifnum\@tempcnta=\@tempcntb \else \def\@citea{--}\fi
    \advance\@tempcnta\m@ne\the\@tempcnta\@citea\the\@tempcntb}\fi\fi}
 
\makeatother
\begin{titlepage}
\pagenumbering{roman}
\CERNpreprint{\DpPaperGroup}{\DpPaperRef} 
\date{{\small\DpDate}} 
\title{\DpTitle} 
\address{\DpAuthors} 
\begin{shortabs} 
\noindent
%
\noindent

A sample of 2.2 million hadronic \z decays, selected from the data recorded by
the {\sc Delphi} detector at {\sc Lep} during 1994-1995 was used for an
improved
measurement of inclusive distributions of \pie, \kp and \p and their
antiparticles in gluon and quark jets. The production spectra of the individual
identified particles were found to be softer in gluon jets compared to quark
jets, with a higher multiplicity in gluon jets as observed for inclusive
charged particles. A significant proton enhancement in gluon jets is observed
indicating that baryon production proceeds directly from colour objects.
The maxima, $\xi^*$, of the $\xi$-distributions for kaons in gluon and quark
jets are observed to be different. 

\end{shortabs}
\vfill
\begin{center}
\DpSubmit \ \\ 
\DpComment \ \\
\DpEMail \ \\
\end{center}
\vfill
\clearpage
\headsep 10.0pt
\addtolength{\textheight}{10mm}
\addtolength{\footskip}{-5mm}
\begingroup
%
\newcommand{\DpName}[2]{\hbox{#1$^{\ref{#2}}$},\hfill}
\newcommand{\DpNameTwo}[3]{\hbox{#1$^{\ref{#2},\ref{#3}}$},\hfill}
\newcommand{\DpNameThree}[4]{\hbox{#1$^{\ref{#2},\ref{#3},\ref{#4}}$},\hfill}
\newskip\Bigfill \Bigfill = 0pt plus 1000fill
\newcommand{\DpNameLast}[2]{\hbox{#1$^{\ref{#2}}$}\hspace{\Bigfill}}
%
\footnotesize
\noindent
\DpName{P.Abreu}{LIP}
\DpName{W.Adam}{VIENNA}
\DpName{T.Adye}{RAL}
\DpName{P.Adzic}{DEMOKRITOS}
\DpName{Z.Albrecht}{KARLSRUHE}
\DpName{T.Alderweireld}{AIM}
\DpName{G.D.Alekseev}{JINR}
\DpName{R.Alemany}{VALENCIA}
\DpName{T.Allmendinger}{KARLSRUHE}
\DpName{P.P.Allport}{LIVERPOOL}
% ...
\normalsize
\endgroup
\titlefoot{Department of Physics and Astronomy, Iowa State
     University, Ames IA 50011-3160, USA
    \label{AMES}}
\titlefoot{Physics Department, Univ. Instelling Antwerpen,
     Universiteitsplein 1, B-2610 Antwerpen, Belgium \\
     \indent~~and IIHE, ULB-VUB,
     Pleinlaan 2, B-1050 Brussels, Belgium \\
     \indent~~and Facult\'e des Sciences,
     Univ. de l'Etat Mons, Av. Maistriau 19, B-7000 Mons, Belgium
    \label{AIM}}
\titlefoot{Physics Laboratory, University of Athens, Solonos Str.
     104, GR-10680 Athens, Greece
    \label{ATHENS}}
\titlefoot{Department of Physics, University of Bergen,
     All\'egaten 55, NO-5007 Bergen, Norway
    \label{BERGEN}}
\titlefoot{Dipartimento di Fisica, Universit\`a di Bologna and INFN,
     Via Irnerio 46, IT-40126 Bologna, Italy
    \label{BOLOGNA}}
\titlefoot{Centro Brasileiro de Pesquisas F\'{\i}sicas, rua Xavier Sigaud 150,
     BR-22290 Rio de Janeiro, Brazil \\
     \indent~~and Depto. de F\'{\i}sica, Pont. Univ. Cat\'olica,
     C.P. 38071 BR-22453 Rio de Janeiro, Brazil \\
     \indent~~and Inst. de F\'{\i}sica, Univ. Estadual do Rio de Janeiro,
     rua S\~{a}o Francisco Xavier 524, Rio de Janeiro, Brazil
    \label{BRASIL}}
\titlefoot{Comenius University, Faculty of Mathematics and Physics,
     Mlynska Dolina, SK-84215 Bratislava, Slovakia
    \label{BRATISLAVA}}
\titlefoot{Coll\`ege de France, Lab. de Physique Corpusculaire, IN2P3-CNRS,
     FR-75231 Paris Cedex 05, France
    \label{CDF}}
\titlefoot{CERN, CH-1211 Geneva 23, Switzerland
    \label{CERN}}
\titlefoot{Institut de Recherches Subatomiques, IN2P3 - CNRS/ULP - BP20,
     FR-67037 Strasbourg Cedex, France
    \label{CRN}}
\titlefoot{Now at DESY-Zeuthen, Platanenallee 6, D-15735 Zeuthen, Germany
    \label{DESY}}
\titlefoot{Institute of Nuclear Physics, N.C.S.R. Demokritos,
     P.O. Box 60228, GR-15310 Athens, Greece
    \label{DEMOKRITOS}}
\titlefoot{FZU, Inst. of Phys. of the C.A.S. High Energy Physics Division,
     Na Slovance 2, CZ-180 40, Praha 8, Czech Republic
    \label{FZU}}
\titlefoot{Dipartimento di Fisica, Universit\`a di Genova and INFN,
     Via Dodecaneso 33, IT-16146 Genova, Italy
    \label{GENOVA}}
\titlefoot{Institut des Sciences Nucl\'eaires, IN2P3-CNRS, Universit\'e
     de Grenoble 1, FR-38026 Grenoble Cedex, France
    \label{GRENOBLE}}
\titlefoot{Helsinki Institute of Physics, HIP,
     P.O. Box 9, FI-00014 Helsinki, Finland
    \label{HELSINKI}}
\titlefoot{Joint Institute for Nuclear Research, Dubna, Head Post
     Office, P.O. Box 79, RU-101 000 Moscow, Russian Federation
    \label{JINR}}
\titlefoot{Institut f\"ur Experimentelle Kernphysik,
     Universit\"at Karlsruhe, Postfach 6980, DE-76128 Karlsruhe,
     Germany
    \label{KARLSRUHE}}
\titlefoot{Institute of Nuclear Physics and University of Mining and Metalurgy,
     Ul. Kawiory 26a, PL-30055 Krakow, Poland
    \label{KRAKOW}}
\titlefoot{Universit\'e de Paris-Sud, Lab. de l'Acc\'el\'erateur
     Lin\'eaire, IN2P3-CNRS, B\^{a}t. 200, FR-91405 Orsay Cedex, France
    \label{LAL}}
\titlefoot{School of Physics and Chemistry, University of Lancaster,
     Lancaster LA1 4YB, UK
    \label{LANCASTER}}
\titlefoot{LIP, IST, FCUL - Av. Elias Garcia, 14-$1^{o}$,
     PT-1000 Lisboa Codex, Portugal
    \label{LIP}}
\titlefoot{Department of Physics, University of Liverpool, P.O.
     Box 147, Liverpool L69 3BX, UK
    \label{LIVERPOOL}}
\titlefoot{LPNHE, IN2P3-CNRS, Univ.~Paris VI et VII, Tour 33 (RdC),
     4 place Jussieu, FR-75252 Paris Cedex 05, France
    \label{LPNHE}}
\titlefoot{Department of Physics, University of Lund,
     S\"olvegatan 14, SE-223 63 Lund, Sweden
    \label{LUND}}
\titlefoot{Universit\'e Claude Bernard de Lyon, IPNL, IN2P3-CNRS,
     FR-69622 Villeurbanne Cedex, France
    \label{LYON}}
\titlefoot{Univ. d'Aix - Marseille II - CPP, IN2P3-CNRS,
     FR-13288 Marseille Cedex 09, France
    \label{MARSEILLE}}
\titlefoot{Dipartimento di Fisica, Universit\`a di Milano and INFN-MILANO,
     Via Celoria 16, IT-20133 Milan, Italy
    \label{MILANO}}
\titlefoot{Dipartimento di Fisica, Univ. di Milano-Bicocca and
     INFN-MILANO, Piazza delle Scienze 2, IT-20126 Milan, Italy
    \label{MILANO2}}
\titlefoot{Niels Bohr Institute, Blegdamsvej 17,
     DK-2100 Copenhagen {\O}, Denmark
    \label{NBI}}
\titlefoot{IPNP of MFF, Charles Univ., Areal MFF,
     V Holesovickach 2, CZ-180 00, Praha 8, Czech Republic
    \label{NC}}
\titlefoot{NIKHEF, Postbus 41882, NL-1009 DB
     Amsterdam, The Netherlands
    \label{NIKHEF}}
\titlefoot{National Technical University, Physics Department,
     Zografou Campus, GR-15773 Athens, Greece
    \label{NTU-ATHENS}}
\titlefoot{Physics Department, University of Oslo, Blindern,
     NO-1000 Oslo 3, Norway
    \label{OSLO}}
\titlefoot{Dpto. Fisica, Univ. Oviedo, Avda. Calvo Sotelo
     s/n, ES-33007 Oviedo, Spain
    \label{OVIEDO}}
\titlefoot{Department of Physics, University of Oxford,
     Keble Road, Oxford OX1 3RH, UK
    \label{OXFORD}}
\titlefoot{Dipartimento di Fisica, Universit\`a di Padova and
     INFN, Via Marzolo 8, IT-35131 Padua, Italy
    \label{PADOVA}}
\titlefoot{Rutherford Appleton Laboratory, Chilton, Didcot
     OX11 OQX, UK
    \label{RAL}}
\titlefoot{Dipartimento di Fisica, Universit\`a di Roma II and
     INFN, Tor Vergata, IT-00173 Rome, Italy
    \label{ROMA2}}
\titlefoot{Dipartimento di Fisica, Universit\`a di Roma III and
     INFN, Via della Vasca Navale 84, IT-00146 Rome, Italy
    \label{ROMA3}}
\titlefoot{DAPNIA/Service de Physique des Particules,
     CEA-Saclay, FR-91191 Gif-sur-Yvette Cedex, France
    \label{SACLAY}}
\titlefoot{Instituto de Fisica de Cantabria (CSIC-UC), Avda.
     los Castros s/n, ES-39006 Santander, Spain
    \label{SANTANDER}}
\titlefoot{Dipartimento di Fisica, Universit\`a degli Studi di Roma
     La Sapienza, Piazzale Aldo Moro 2, IT-00185 Rome, Italy
    \label{SAPIENZA}}
\titlefoot{Inst. for High Energy Physics, Serpukov
     P.O. Box 35, Protvino, (Moscow Region), Russian Federation
    \label{SERPUKHOV}}
\titlefoot{J. Stefan Institute, Jamova 39, SI-1000 Ljubljana, Slovenia
     and Laboratory for Astroparticle Physics,\\
     \indent~~Nova Gorica Polytechnic, Kostanjeviska 16a, SI-5000 Nova Gorica, Slovenia, \\
     \indent~~and Department of Physics, University of Ljubljana,
     SI-1000 Ljubljana, Slovenia
    \label{SLOVENIJA}}
\titlefoot{Fysikum, Stockholm University,
     Box 6730, SE-113 85 Stockholm, Sweden
    \label{STOCKHOLM}}
\titlefoot{Dipartimento di Fisica Sperimentale, Universit\`a di
     Torino and INFN, Via P. Giuria 1, IT-10125 Turin, Italy
    \label{TORINO}}
\titlefoot{Dipartimento di Fisica, Universit\`a di Trieste and
     INFN, Via A. Valerio 2, IT-34127 Trieste, Italy \\
     \indent~~and Istituto di Fisica, Universit\`a di Udine,
     IT-33100 Udine, Italy
    \label{TU}}
\titlefoot{Univ. Federal do Rio de Janeiro, C.P. 68528
     Cidade Univ., Ilha do Fund\~ao
     BR-21945-970 Rio de Janeiro, Brazil
    \label{UFRJ}}
\titlefoot{Department of Radiation Sciences, University of
     Uppsala, P.O. Box 535, SE-751 21 Uppsala, Sweden
    \label{UPPSALA}}
\titlefoot{IFIC, Valencia-CSIC, and D.F.A.M.N., U. de Valencia,
     Avda. Dr. Moliner 50, ES-46100 Burjassot (Valencia), Spain
    \label{VALENCIA}}
\titlefoot{Institut f\"ur Hochenergiephysik, \"Osterr. Akad.
     d. Wissensch., Nikolsdorfergasse 18, AT-1050 Vienna, Austria
    \label{VIENNA}}
\titlefoot{Inst. Nuclear Studies and University of Warsaw, Ul.
     Hoza 69, PL-00681 Warsaw, Poland
    \label{WARSZAWA}}
\titlefoot{Fachbereich Physik, University of Wuppertal, Postfach
     100 127, DE-42097 Wuppertal, Germany
    \label{WUPPERTAL}}
\addtolength{\textheight}{-10mm}
\addtolength{\footskip}{5mm}
\clearpage
\headsep 30.0pt
\end{titlepage}
%
\pagenumbering{arabic} 
\setcounter{footnote}{0} %
\large
%
\section{Introduction}

The different colour charges of quarks and gluons lead to specific differences
in the particle multiplicity, the energy spectrum and the angular distributions
of the corresponding jets. Beyond the study of these
differences~\cite{mulkhoze}, which are
related to the perturbative properties of
QCD\footnote{{\bf Q}uantum{\bf C}hromo{\bf D}ynamics} elementary fields, the
comparison of gluon and quark jets opens up the possibility to
infer properties of the non-perturbative formation of hadrons directly.

The study of ratios of identified ($\pi^\pm, K^\pm, p(\bar{p}))$ particle
distributions in gluon ($g$) and quark ($q$) jets is the main
subject of this paper.

Gluon jets are selected in \bbg events by tagging the \b quarks using
techniques based on the large impact parameters of tracks coming from heavy
particle decays. The Ring Imaging Cherenkov Counters (RICH) of the {\sc Delphi}
detector  provide particle identification over a wide momentum range in
combination with the ionization loss measurement of the Time Projection Chamber
(TPC) and so allow a detailed comparison of identified particle spectra in
gluon and quark jets. These are used for a detailed test of QCD based
fragmentation models and also to check
MLLA\footnote{{\bf M}odified {\bf L}eading {\bf L}og {\bf A}pproximation}
and LPHD\footnote{{\bf L}ocal {\bf P}arton {\bf H}adron {\bf D}uality}
predictions~\cite{greybook}.

This paper is organized as follows. In Section~\ref{sec_exp} the hadronic
event selection, the quark/gluon separation, and the particle identification
are described briefly. The experimental
results are presented and
compared with the predictions of models in Section~\ref{sec_res}. Finally
a summary and  conclusions are presented in Section~\ref{sec_sum}.

\section{Experimental Technique and Event Sample}
\label{sec_exp}

A description of the {\sc Delphi} detector, together
with a description of its performance, can be found in~\cite{perfo}.

\subsection{Event Selections}
\label{sec_event}

The data collected by {\sc Delphi} during 1994-1995 are
considered in the present analysis,
during which time the RICH~\cite{perfo} detectors
(the main particle identification detectors) were fully operational and the
Vertex detector was equipped with a three-dimensional readout.
The cuts applied to charged and neutral particles and to events in order
to select hadronic Z decays are identical to those given
in~\cite{splitting_paper} and~\cite{disboth}.
The data sample passing the selection of hadronic events contained  1,775,230
events with a small contamination  ($<0.7\%$) arising from
$\tau^+\tau^-$  pairs, beam-gas scattering and $\gamma\gamma$
interactions~\cite{perfo}.

The influence of the detector performance
on the analysis was studied with the full {\sc
Delphi} simulation program, {\sc Delsim}~\cite{perfo}. Events generated with
the {\sc Jetset} 7.3 Parton Shower (PS) model~\cite{lun1}, with parameters
tuned by {\sc Delphi}~\cite{hamacher}, were passed through {\sc Delsim} and
processed with the same reconstruction and analysis programs as the real data.

Three-jet events were clustered using the
Durham algorithm~\cite{dur1} with a jet resolution parameter $y_{cut}=0.015$.
The value used for the cut-off was optimized using the
{\sc Jetset} 7.3 PS model, by maximizing the statistics available
and the quark/gluon purity attained for the three-jet event
samples~\cite{fuster}.

\begin{figure}[t,b,h]
\centering\parbox{\bwid}{
\centering\includegraphics[width=\fwid]{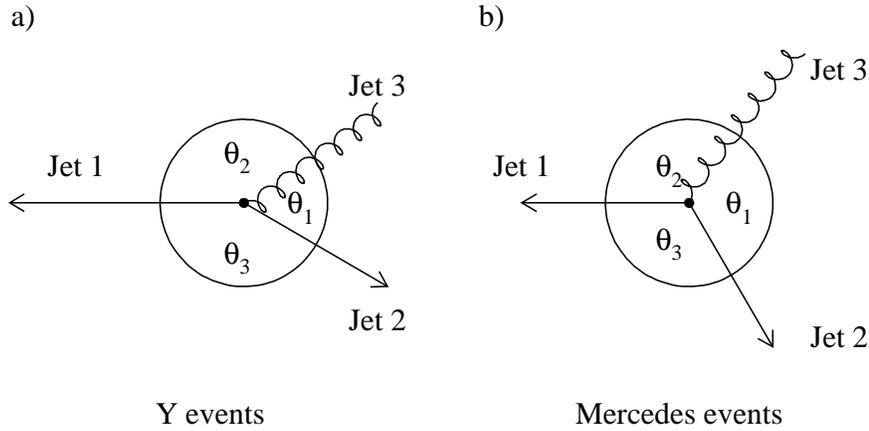}
\caption{\label{qcd_1_s}Event topologies of symmetric Y events and
Mercedes events; $\theta_{i}$ are the angles between the jets after projection
into the event plane.}
}
\end{figure}

The jet axes were projected onto the event plane, defined
as the plane perpendicular to the smallest sphericity eigenvector  obtained
from the quadratic momentum
tensor, $M_{\alpha\beta}=\sum_{i=1}^n p_{i\alpha}p_{i\beta}$.
The jets were numbered in decreasing order of jet energy, where the energy
of each jet is calculated from the angles between the jets assuming  massless
kinematics:
\begin{eqnarray}
\label{recalibration}
E_{j}\sp{calc} =   { {\rm sin} \theta_{j}   \over
                    {\rm sin} \theta_{1} +
                    {\rm sin} \theta_{2} +
                    {\rm sin} \theta_{3}  } \sqrt{s},
                    \ \ \ \ j=1,2,3\, ,
\end{eqnarray}
where $\theta_{j}$  is the interjet angle as defined in~\fref{qcd_1_s}.

For a detailed comparison of quark and  gluon jet
properties,  it is necessary to obtain samples of quark  and gluon jets with
similar kinematics and the same underlying scales~\cite{scaling}.
To fulfill  this condition,
two different event topologies were used,
as  illustrated in~\fref{qcd_1_s}:
\begin{itemize}
 \item mirror symmetric events,
       with $\theta_{2}$ and $\theta_{3}
       \in [150^\circ-15^\circ,150^\circ+15^\circ]$,\\
       subsequently called {\bf Y events}, and
 \item three-fold symmetric events,
       with $\theta_{2}$ and $\theta_{3}
       \in [120^\circ-15^\circ,120^\circ+15^\circ]$,\\
       subsequently called {\bf Mercedes events}.
\end{itemize}

\bwid 0.8\textwidth
\fwid 0.99\bwid
For Y events only the low energy jets (jets 2
and 3 in~\fref{qcd_1_s}) were used in the analysis. For Mercedes
events all jets were used in the analysis.
The appropriate scale for these jets, equivalent to the
$e^+e^-$ beam energy can be approximated by 
$\kappa=E_{jet} \sin\theta_1/2$~\cite{scaling}.
Mercedes events are mainly used to study the scale dependence of particle 
production.

In order to enhance the contribution from events with three well-defined jets
attributed to $q\bar{q}g$ production, further cuts 
(sum of angles between jet, polar angle of each jet axis, visible jet energy
per jet and number of particles in each jet)
are applied to the
three-jet event samples, as in~\cite{splitting_paper}.
The number of three-jet events in the Mercedes and Y samples is 
11,685 and 110,628 respectively.

~

~

\subsection{Quark and Gluon Jet Identification}
\label{technics}

The identification of gluon jets by anti-tagging of heavy quark jets
is identical to that described in~\cite{splitting_paper}.
Heavy quark tagging is based on large impact parameters with respect to the
primary vertex due to the long lifetime of the heavy particles.

The efficiency and purity calculations were made using events generated by the
{\sc Jetset} 7.3 Monte Carlo model tuned to {\sc Delphi} data \cite{hamacher} and
passed through {\sc
Delsim}.

Even in simulated events, the assignment of parton flavours to the jets is not
unique, as the decay history is interrupted
by the building of strings in models such as {\sc Jetset}
or by the parton assignment of clusters in the case of {\sc Herwig}.
Thus two independent
ways  of defining the gluon jet in the fully simulated events were
investigated. The first method
assumed that the jet which
has the  largest angle to hadrons containing heavy quarks is
the gluon induced jet~\footnote{There are almost always only two heavy
hadrons in an event, because the $g \to q\bar{q}$ splitting into heavy quarks
is strongly suppressed.}  (angle assignment) and in the second method
the jet
containing the fewest decay particles from the heavy hadrons was
assigned to the
gluon (history assignment).
Both methods give similar results and
therefore the purities can be estimated with small systematic  
uncertainties~\cite{disboth}.

Gluon jet purities of
$\sim 82\%$ for Y events and Mercedes events were achieved.
Here the purity is defined as the ratio of correctly identified
gluon jets to the total number
of jets tagged as gluons.
There are 24,449 events with an identified gluon jet in the case of Y events and
1,806 in the case of Mercedes events.

\subsubsection{Corrections}
\label{nmix}
\begin{table}[b,t,h]
\begin{center}

\begin{tabular}{|c|c|c|c|}\hline
 {\rul \bf Y}     &  {\rul \bf quark content ($dusc$)} &
 {\rul \bf $b$ quark content} & {\rul \bf gluon content} \\
\hline
\hline
 normal mixture    & 49.5\%  &   ~1.6\% & 48.9\% \\
\hline
 $b$ tagged jets     & 25.1\%  & 58.2\% & 16.6\% \\
\hline
 gluon tagged jets & 13.7\%  &   ~4.2\% & 82.0\% \\
\hline
\hline
 {\rul \bf Mercedes}     &  {\rul \bf quark content ($dusc$)} &
 {\rul \bf $b$ quark content} & {\rul \bf gluon content} \\
\hline
\hline
 normal mixture    & 64.2\%  &   ~2.3\% & 33.3\% \\
\hline
 $b$ tagged jets     & 17.3\%  & 73.6\% & ~9.1\% \\
\hline
 gluon tagged jets & 11.1\%  &   ~6.8\% & 81.8\% \\
\hline
\end{tabular}
\caption{Compositions of different jet classes in Y and Mercedes
events. The statistical
errors are smaller than 1\%.}
\label{tb_matrix2}

\end{center}
\end{table}

\tref{tb_matrix2} shows the fractions of ``light'' quark, $b$ quark and gluon
jets in the three different jet classes entering the analysis of
Y and Mercedes events.
The classes are normal mixture jets, gluon tagged jets and b-tagged jets. 
``Light'' quark denotes here a mixture of $dus$ and $c$ quarks. 
The jets of the normal mixture
are taken from events in which the heavy hadron tag failed.
Therefore they are
predominantly unidentified $dusc$ quark and gluon jets.
Denoting a data bin of an 
observable of a pure gluon, light or $b$ quark jet sample with $R_g$,
$R_{dusc}$ and $R_b$ respectively, the measured observables in the three
tagged classes $R_{g_{tag}}$, $R_{b_{tag}}$ and $R_{mix}$ can be written as:
\begin{eqnarray}
 \nonumber
 R_{mix}   &=& p_{mix}^{dusc} \cdot R_{dusc} + p_{mix}^{b} \cdot R_{b}
 + p_{mix}^g \cdot R_g  \\
 \label{gl_pur}
 R_{b_{tag}}   &=& p_{b_{tag}}^{dusc} \cdot R_{dusc} + p_{b_{tag}}^{b} \cdot R_{b}
 + p_{b_{tag}}^g \cdot R_g \\
 \nonumber
 R_{g_{tag}}   &=& p_{g_{tag}}^{dusc} \cdot R_{dusc} + p_{g_{tag}}^{b} \cdot R_{b}
 + p_{g_{tag}}^g \cdot R_g
\end{eqnarray}
where the $p_i^j$ are the fractions as e.g. shown in~\tref{tb_matrix2}. 
The observables
for the pure samples can then be obtained by solving equation \ref{gl_pur} for
$R_g$, $R_l$ and $R_b$. 
The statistical errors on the fractions
$p_i^j$ are less than $1\%$ and are fully propagated 
with only a small effect on the total errors.
Instead of the ``light'' quark sample containing $dus$ and $c$ quarks, a
general quark sample containing $dusc$ and $b$ quarks can be deduced
by setting $p_{b_{tag}}^x=0$ and adding $p^b_{mix}$ to $p^{dusc}_{mix}$
and $p^b_{g_{tag}}$ to $p^{dusc}_{g_{tag}}$, reducing equation 2 to a
$2 \times 2$ matrix equation. In a similar way  the light quark sample
can be reduced to only containing $dus$ but no $c$ quarks by hardening
the heavy hadron tag and treating the $c$ quarks like $b$ quarks in
the Monte-Carlo. This leads to the pure observables $R_g$, $R_{dus}$
and $R_{cb}$ which are related to the measured observables
$R_{g_{tag}}$, $R_{b_{tag}}$ and $R_{mix}$ in the same way as before
but with different $p_i^j$.
To correct for the limited detector acceptance, secondary
reinteraction of particles and resolution of the detector, an acceptance
correction factor 
\begin{equation}
 C^{acc} = \frac{R^{MC}}{R^{MC+detector}}
\label{e:accor}
\end{equation}
is also applied to the data bin by bin for each distribution. Here
$R^{MC}$ denotes pure model distributions (referring to ``light'' quark and
gluon jets) and $R^{MC+detector}$ denotes the
full simulation including detector effects treated like the data. 
Long lived particles like
the $K^0$ and the $\Lambda^0$ were considered as instable when
computing model distributions.

\subsection{Identification of Final State Particles}
\label{sec_partid}

For the measurement of the \pie, \kp and proton content in jets
a combined tagging procedure
based on the Cherenkov angle measurement in the RICH detector and on the
ionization energy loss ($dE/dx$) in the TPC was applied
which is described in detail in~\cite{perfo}.

The combined application of TPC and RICH allows a continuous particle
identification in the momentum range of 0.3-45.0~${\rm GeV/c}$.
\tref{t:combined} shows which detectors were used
to identify pions, kaons, and protons depending on their momentum.

\begin{table}[b,h,t]
\begin{center}
\begin{tabular}{|c|c|c|c|c|c|c|c|}
   \hline
   & \mc{7}{c|}{Momentum Range [$GeV/c$]}\\
  \cline{2-8}
   & 0.3 - 0.7 & 0.7 - 0.9 & 0.9 - 1.3 & 1.3 - 2.7 & 2.7 - 9.0 &
                                         9.0 -16.0 & 16.0 - 45.0 \\
  \hline
  \hline
       &     & \mc{3}{c|}{}        & \mc{3}{c|}{}\\
 $\pi$ & TPC & \mc{3}{c|}{LRICH S} & \mc{3}{c|}{GRICH S}\\
       &     & \mc{3}{c|}{}        & \mc{3}{c|}{}\\
  \hline
       &     & \mc{3}{c|}{}        & GRICH V &\mc{2}{c|}{}\\
   K   & TPC & \mc{3}{c|}{LRICH S} &    +    &\mc{2}{c|}{GRICH S}\\
       &     & \mc{3}{c|}{}        & LRICH S &\mc{2}{c|}{}\\
  \hline
       & \mc{2}{c|}{}    &  TPC    &         & GRICH V &         &         \\
   p   & \mc{2}{c|}{TPC} &  +      & LRICH S &    +    & GRICH V & GRICH S \\
       & \mc{2}{c|}{}    & LRICH V &         & LRICH S &         &         \\
  \hline
\end{tabular}

\begin{tabular}{ll}
   TPC        & Identification by measurement of the energy loss\\
   LRICH S(V) & Signal (Veto)-Identification with the liquid RICH\\
   GRICH S(V) & Signal (Veto)-Identification with the gas RICH\\
\end{tabular}
\caption{Application ranges of the detectors for particle identification}{
\vspace{0.8cm}
}
\label{t:combined}
\end{center}
\vspace{-1cm}
\end{table}

An algorithm was developed
to obtain an optimal combination of the particle identification
possibilities of the TPC and the RICH. It combines
the probabilities for the particle identification with the TPC and the RICH
by a simple multiplication and renormalization, and predefines three 
different identification classes,
{\em loose}, {\em standard}, and {\em tight}, by using well chosen cuts on this
combined probability distribution.
These cuts for the particle identification probabilities
allow particle identification performances with different purities $R$ and
efficiencies $\eps$:

\begin{eqnarray*}
R_i^j    &=& \frac{ \m{\# of particles of kind $i$ identified as kind $j$}}
                  { \m{\# of all particles identified as kind $j$ }}, \\[0.5ex] 
\eps_i^j &=& \frac{ \m{\# of particles of kind $i$ identified as kind $j$}}
                  { \m{\# of all particles of kind $i$ }}\  .\\
\end{eqnarray*}

\bwid \textwidth \fwid 0.69\bwid
\begin{figure}[p]
\centering\parbox{\bwid}{
\centering\includegraphics[height=0.75\textheight]{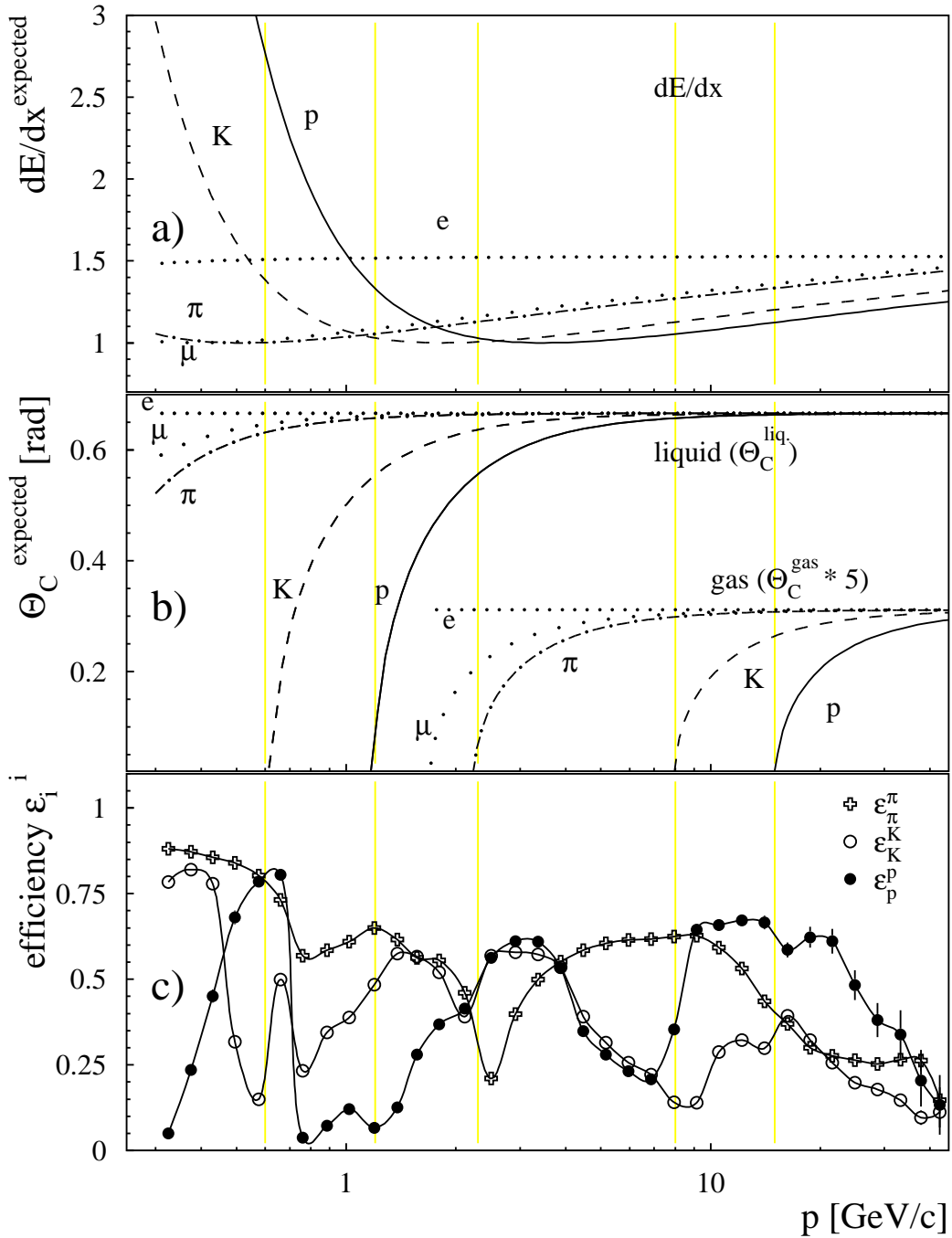}
\caption{Curves of expected values and efficiencies of the particle
identification;
a) shows the curve of the expected values for specific ionization
for pions ($\pi$), kaons ($K$), protons ($p$), muons ($\mu$),
and electrons ($e$) as a function of the momentum.
b) shows the curve of the expected values for the Cherenkov angle $\thetac$
in the liquid and gas radiator for the same particle hypotheses.
$\thetac^{Gas}$ was multiplied by a factor 5.
The curves begin at $p=0.3\gevc$ for the liquid radiator and at $1.7\gevc$
for the gas radiator.
c) shows the resulting efficiencies for Y events
for the standard identification
of pions, kaons and protons in the barrel of {\sc Delphi} for the 1994-95 data.
Light vertical lines in all plots indicate the threshold-momenta
of $\pi$, $K$ and $p$ identification in the two RICH radiators.
}
\label{f:pa-erwart}
}
\end{figure}

\fref{f:pa-erwart} shows the efficiency of the combined particle identification
of pions, kaons, and protons as a function of the momentum of the particle.
The curves of the expected energy loss and the Cherenkov angle, \thetac,
are shown in the upper part of
\fref{f:pa-erwart}.

\begin{figure}[p]
 \rotatebox{90}{
  \bwid 0.75\textheight
  \fwid 0.73\textheight
  \parbox[t]{\bwid}{
   \begin{center}
    \includegraphics[width=\fwid]{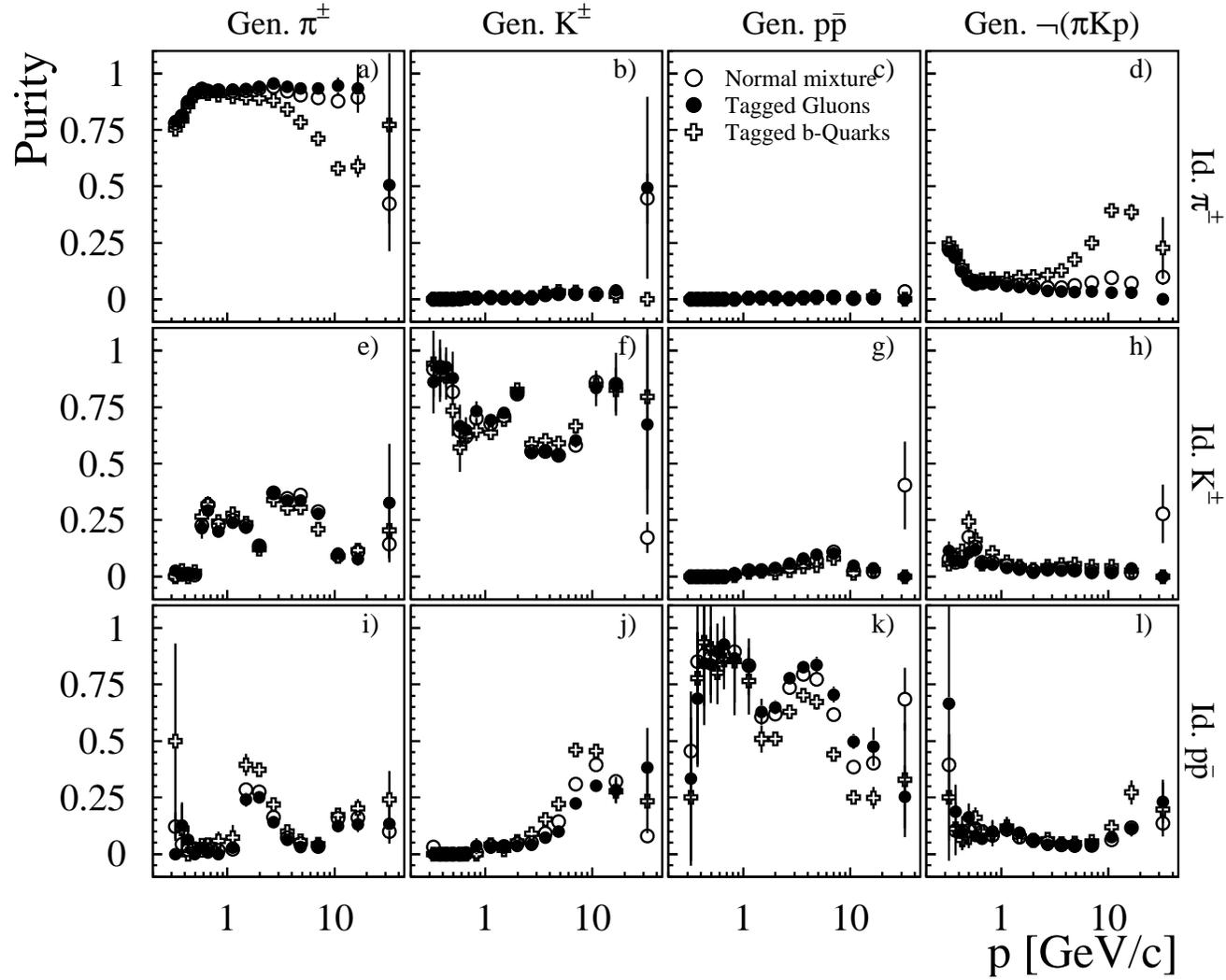}
    \caption{Purity of the particle identification in Y events.
             Here `Gen.' denotes the
             generated flavour of the particle and `Id.' denotes the
             tagged particle flavour. $\neg$ means not.}
    \label{f:purinopos}
   \end{center}
   }
 }
\end{figure}

\fref{f:purinopos} shows the resulting purities of the particle identification
for Y events.
To keep the influence of particle reinteractions in the detector material
small, this distribution is restricted to negatively
charged particles in the momentum range $p<2.7$\gevc
(for details see~\cite{disboth}). 
The purity matrix is predominantly diagonal (\fref{f:purinopos}a,f,k).
The most important, however still negligible background for the pion
reconstruction stems from electrons and muons.
An exception are energetic electrons and muons
from semileptonic hadron decays in $b$ (and $c$) jets
with an identification rate up to 20\%. The main background for the kaon
selection are pions. A kaon identification purity of $\sim 70\%$ is achieved.
As the kaon production rate is almost one order of magnitude smaller than the
pion production rate in hadronic
Z decays, this implies a very efficient pion suppression. The lower
proton 
identification purity in $b$ jets is mainly
due to the higher probability to identify kaons
as protons because of the higher K multiplicity in $b$ jets.

\subsubsection*{Acceptance correction of the spectra of identified particles}

From the measured particle spectra $I_\pi, I_K$, and $I_p$ of identified
particles one obtains the spectra of pure hadrons $S_\pi, S_K$, and $S_p$
by solving the equation system:
\begin{eqnarray}
\renewcommand{\arraystretch}{1.3}
\left(\begin{array}{c}I_\pi\\I_K\\I_p\end{array}\right) =
\left(\begin{array}{ccc}\eps^\pi_\pi&\eps^\pi_K&\eps^\pi_p\\
              \eps^K_\pi&\eps^K_K&\eps^K_p\\
              \eps^p_\pi&\eps^p_K&\eps^p_p\end{array}\right) \cdot
\left(\begin{array}{c}S_\pi\\S_K\\S_p\end{array}\right)\ .
\end{eqnarray}
This correction is applied before the correction of the jet purity.
The values $\eps_i^i$ denote the efficiencies that the particles $i$ are identified
correctly; the values $\eps_i^j$ with $i\neq j$ are proportional to the
background of particle class $j$.
A correction for secondary reinteraction of particles in the
detector material was included in the overall correction factor Equation 
\ref{e:accor}.
\section{Results}
\label{sec_res}

For identified particles in quark and gluon jets the multiplicty and
the semi-inclusive distributions as function of the momentum $p$, 
$\xi_p=\ln1/x_p=\ln p_{jet}/p$ 
and the rapidity $\eta=\ln\frac{E+p_{\parallel}}{E-p_{\parallel}}$
with respect to the jet axis have been measured.
For each particle also the Ratio between gluon and quark 
jets $R$ and the normalized ratio $R'=R/r_{ch}$ of each of the
observables have been studied, where $r_{ch}$ denotes the corresponding ratio
obtained for all charged particles.


Note that the particle multiplicity of jets is not a well
defined subject which depends on details of the jet definition influencing the
assignment of low momentum particles to the jets. The given results on
multiplicities and also the particle distribution corresponding to very small
momenta therefore always refer to the jet definition specified in
Section~\ref{sec_event}.


Special emphasis here lies on the measurement of ratios in gluon to quark jets
$R$ as in these ratios the systematic error is considerably reduced as
most of the systematic uncertainties cancel out. The double ratios
$R^{'}$ stress particle specific differences between gluon and quark jets.

In this analysis gluon jets are in general compared to a "duscb" quark jet
reference sample. The flavour mix of this sample 
is that of hadronic Z decays. For the comparison of particle
multiplicities also reference samples were used where the b events
("dusc") and all heavy quark events ("dus") were removed.

\subsection{Multiplicities}

\begin{figure}[htb]
\centering\parbox{\bwid}{
\centering\includegraphics[width=\fwid]{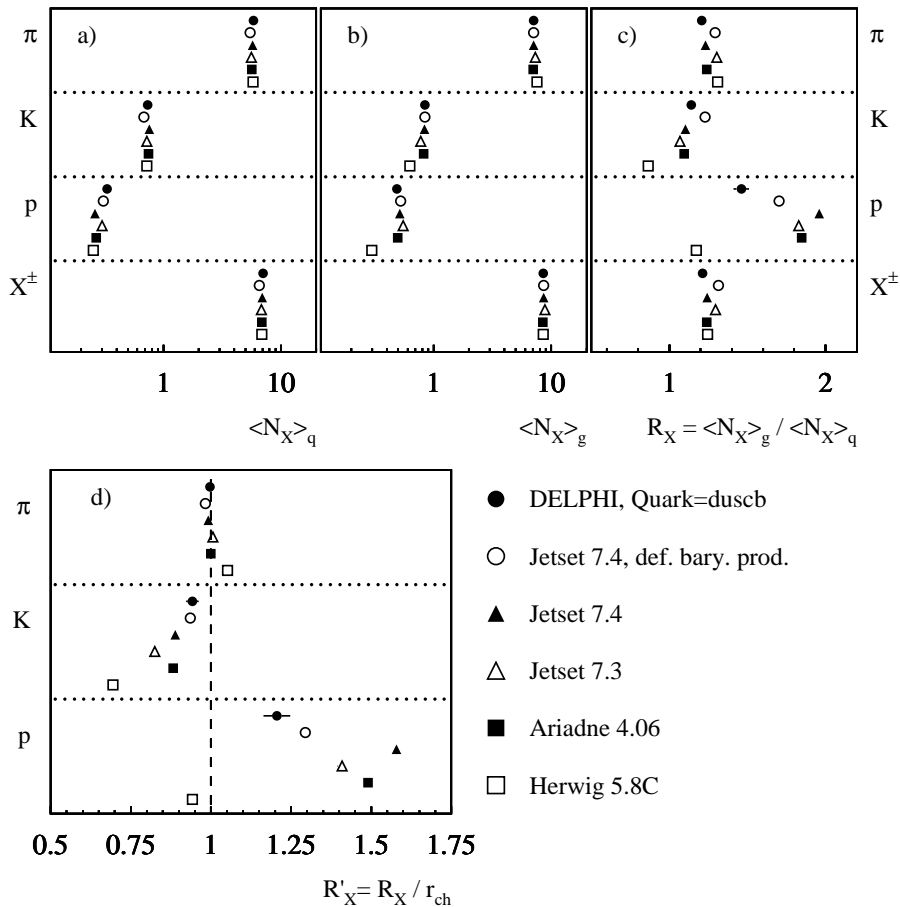}
\caption{Mean multiplicities $N_X$and ratios of multiplicities $R_X$
         for identified particles $X$ in quark and gluon jets
         of Y events compared to different Monte Carlo models.}
\label{f:pa-NUM-GENMC}
}
\end{figure}

In~\fref{f:pa-NUM-GENMC} we present the mean multiplicities $N_q$ and $N_g$
for identified particles in quark and gluon jets respectively,
as well as their ratio,
$R=N_g/N_q$,  and the normalized multiplicity ratios, $R'=R/r_{ch}$. 

\bwid 0.99\textwidth \fwid 0.8\bwid

The multiplicities measured in quark jets for identified hadrons and for
all charged hadrons depend on the composition of the quark flavours within the
quark jet sample. 
The values obtained for the multiplicities $N_g$  and $N_q$ and the
ratios $R$ resp. $R'$ are given in~\tref{t:part:erg}. The
determination of the systematic errors is described below.
In~\tref{tb_rat} the normalized multiplicity ratios $R'$ are compared to
the predictions from the Monte Carlo simulations for Y and Mercedes events. The
data show a significant proton enhancement in gluon jets for Y events.
A similar enhancement, although less significant, is also seen in Mercedes
events.
The slight change observed for $R'_K$ and $R'_p$ (see~\tref{t:part:erg})
for different flavour
compositions can be understood due to a stronger $K$ production and a depleted
proton production in events with heavy quarks.

Simulations with statistics superior to the data
based on the {\sc Jetset} 7.4 PS model, {\sc Ariadne}
4.08~\cite{ari}, and~{\sc Herwig} 5.8~\cite{herw} with parameters tuned by {\sc
Delphi}~\cite{hamacher} are compared to the data.
{\sc Herwig} underestimates both the kaon and the proton production in gluon
jets. In contrast {\sc Jetset} and {\sc Ariadne}~\footnote{
Note that {\sc Ariadne} employs the non-perturbative hadronization model of
{\sc Jetset}.
} tend to overestimate the proton
production in gluon jets. 
The {\sc Jetset} model with default baryon production
\footnote{different from~\cite{hamacher} $mstj(12)=2$} deviates less. 
The difference to the other model is that here the extra suppression at the
string end, which had been introduced to  describe baryon production at large
scaled momenta better~\cite{hamacher}, is inactive.
The excess of baryon production in gluon jets
indicates that baryons are directly produced from a colour string   and not via
intermediate colour    and baryon number neutral clusters. This is discussed in
more detail in Section~\ref{sec_baryon}.

\BT[htb]
\BC
\BTB{|c|c|cc|cc|}\hline
\mc{6}{|c|}{Y events}\\\hline\hline
Kind & Quark & $N_{Quark}$ & $N_{Gluon}$ & $R_X$ & $R'_X$ \\\hline
        &duscb   &\erg{5.852}{0.036}{0.071} & \erg{7.067}{0.032}{0.077} &
                  \erg{1.208}{0.009}{0.020} & \erg{0.997}{0.009}{0.013} \\
$\pi$   &dusc    &\erg{5.702}{0.039}{0.069} & \erg{7.043}{0.031}{0.076} &
                  \erg{1.235}{0.010}{0.020} & \erg{0.977}{0.010}{0.013} \\
        &dus     &\erg{5.672}{0.040}{0.067} & \erg{7.051}{0.032}{0.076} &
                  \erg{1.243}{0.010}{0.020} & \erg{0.966}{0.010}{0.012} \\\hline
        &duscb   &\erg{0.737}{0.012}{0.013} & \erg{0.841}{0.010}{0.015} &
                  \erg{1.141}{0.023}{0.019} & \erg{0.942}{0.019}{0.016} \\
$K$     &dusc    &\erg{0.692}{0.013}{0.013} & \erg{0.836}{0.010}{0.015} &
                  \erg{1.208}{0.026}{0.020} & \erg{0.956}{0.021}{0.018} \\
        &dus     &\erg{0.637}{0.013}{0.012} & \erg{0.835}{0.010}{0.015} &
                  \erg{1.310}{0.031}{0.022} & \erg{1.018}{0.025}{0.018} \\\hline
        &duscb   &\erg{0.332}{0.010}{0.004} & \erg{0.485}{0.009}{0.008} &
                  \erg{1.460}{0.050}{0.031} & \erg{1.205}{0.041}{0.025} \\
$p$     &dusc    &\erg{0.333}{0.010}{0.004} & \erg{0.494}{0.009}{0.008} &
                  \erg{1.481}{0.053}{0.031} & \erg{1.172}{0.043}{0.025} \\
        &dus     &\erg{0.343}{0.011}{0.004} & \erg{0.489}{0.009}{0.008} &
                  \erg{1.427}{0.051}{0.028} & \erg{1.109}{0.040}{0.021} \\\hline
        &duscb   &\erg{7.077}{0.031}{0.071} & \erg{8.573}{0.026}{0.086} &
                  \erg{1.211}{0.006}{0.014} & 1 \\
$X^\pm$ &dusc    &\erg{6.773}{0.033}{0.068} & \erg{8.560}{0.026}{0.086} &
                  \erg{1.264}{0.007}{0.015} & 1 \\
        &dus     &\erg{6.654}{0.034}{0.067} & \erg{8.559}{0.026}{0.086} &
                  \erg{1.286}{0.008}{0.015} & 1 \\\hline\hline
\mc{6}{|c|}{Mercedes events}\\\hline\hline
Kind & Quark & $N_{Quark}$ & $N_{Gluon}$ & $R_X$ & $R'_X$ \\\hline
        &duscb   &\erg{6.973}{0.078}{0.085} & \erg{8.962}{0.133}{0.096} & 
                  \erg{1.285}{0.024}{0.022} & \erg{0.998}{0.023}{0.012} \\
$\pi$   &dusc    &\erg{6.735}{0.084}{0.076} & \erg{9.002}{0.134}{0.133} & 
                  \erg{1.337}{0.026}{0.029} & \erg{0.988}{0.023}{0.012} \\
        &dus     &\erg{6.700}{0.088}{0.072} & \erg{9.028}{0.134}{0.133} & 
                  \erg{1.347}{0.027}{0.029} & \erg{0.974}{0.024}{0.012} \\\hline
        &duscb   &\erg{0.862}{0.025}{0.014} & \erg{0.978}{0.041}{0.016} & 
                  \erg{1.135}{0.058}{0.021} & \erg{0.881}{0.046}{0.018} \\
$K$     &dusc    &\erg{0.819}{0.027}{0.013} & \erg{0.982}{0.042}{0.011} & 
                  \erg{1.199}{0.064}{0.017} & \erg{0.886}{0.049}{0.016} \\
        &dus     &\erg{0.773}{0.027}{0.013} & \erg{0.981}{0.042}{0.011} & 
                  \erg{1.268}{0.070}{0.018} & \erg{0.917}{0.052}{0.017} \\\hline
        &duscb   &\erg{0.401}{0.022}{0.006} & \erg{0.656}{0.040}{0.013} & 
                  \erg{1.635}{0.134}{0.049} & \erg{1.269}{0.106}{0.046} \\
$p$     &dusc    &\erg{0.422}{0.023}{0.012} & \erg{0.636}{0.038}{0.013} & 
                  \erg{1.507}{0.123}{0.098} & \erg{1.114}{0.092}{0.065} \\
        &dus     &\erg{0.408}{0.025}{0.013} & \erg{0.674}{0.041}{0.016} & 
                  \erg{1.650}{0.142}{0.115} & \erg{1.192}{0.104}{0.071} \\\hline
        &duscb   &\erg{8.467}{0.066}{0.098} & \erg{10.91}{0.113}{0.110} & 
                  \erg{1.288}{0.017}{0.018} & 1 \\
$X^\pm$ &dusc    &\erg{8.085}{0.071}{0.097} & \erg{10.94}{0.113}{0.123} & 
                  \erg{1.353}{0.018}{0.025} & 1 \\
        &dus     &\erg{7.942}{0.074}{0.089} & \erg{10.99}{0.114}{0.124} & 
                  \erg{1.384}{0.019}{0.024} & 1 \\\hline
\ETB
\caption{Multiplicities  and ratios of multiplicities
         of identified particles in \qgj.}
\label{t:part:erg}
\EC\ET

\begin{table}[htbp]
\begin{center}
\begin{tabular}{|l||c|c|c|c|c|c|}\hline
 {\rul \bf $R'_X$} & {\rul \bf Data} & 
 {\rul \bf JT 74 def. bary.} &
 {\rul \bf JT 74} &  
 {\rul \bf JT 73} &
 {\rul \bf AR} & {\rul \bf HW} \\
 \hline
\multicolumn{7}{|c|}{{\rul \bf Y Events}}\\ \hline
 $R'_{\pi^+}$   & \erg{0.997}{0.009}{0.013} &0.98&0.99&1.01&1.00&1.05\\
 $R'_{K^+}$     & \erg{0.942}{0.019}{0.016} &0.94&0.89&0.82&0.88&0.69\\
 $R'_{p}$       & \erg{1.205}{0.041}{0.025} &1.29&1.58&1.41&1.49&0.94\\
\hline
\multicolumn{7}{|c|}{{\rul \bf Mercedes Events}}\\ \hline
 $R'_{\pi^+}$   & \erg{0.998}{0.023}{0.012} &1.00&1.01&1.02&1.01&1.05\\
 $R'_{K^+}$     & \erg{0.881}{0.046}{0.018} &0.94&0.90&0.81&0.88&0.70\\
 $R'_{p}$       & \erg{1.269}{0.106}{0.046} &1.20&1.43&1.38&1.37&1.11\\
\hline
\end{tabular}
\end{center}
\caption{
Normalized multiplicity ratios $R'_X$ (for $duscb$ quarks) compared to the
predictions from the Monte Carlo simulations (JT 74 def. bary. =
{\sc Jetset} 7.4 PS with default baryon production, JT 74 = {\sc Jetset} 7.4
PS, JT 73 = {\sc Jetset} 7.3 PS, AR = {\sc Ariadne} 4.08,
HW = {\sc Herwig} 5.8C).}
\label{tb_rat}
\end{table}

As a cross-check the summed multiplicity ratio $ R_{\pi^{\pm}+p^{\pm}+K^{\pm}}$
was calculated. A value of $1.21 \pm 0.01$ was obtained in the case of Y events
and $1.29 \pm 0.02$ in the case of Mercedes events. Both numbers are in
good agreement with a direct measurement of this ratio~\cite{Bagliesi:1998mp}
(Y: $1.235 \pm 0.030$, Mercedes: $1.276 \pm 0.059$).

For completeness \tref{t:rprime_koll} shows a comparison
with measurements of other experiments. A significant excess of proton
production was observed by {\sc Argus}~\cite{l:argus:part} and
{\sc Opal}~\cite{Bagliesi:1998mp}. No quantitative comparison is,
however, possible due to the different energies or event topologies.

\subsection*{Systematic Errors}

\tref{t:partsys-y} summarizes the influences of the most important sources
of systematic error for the determination of the multiplicities  and their
ratios. To obtain systematic errors comparable with the statistical errors,
half the difference of the value obtained when a parameter is modified from its
central value is quoted as the systematic uncertainty. The single errors are
added quadratically. The following sources of systematic uncertainties were
examined.

\begin{enumerate}
\item {\bf Decays of $K^0, \Lambda^0$} \\
      It was examined whether the ratios of the production rates of pions,
      kaons and protons in quark and gluon jets are influenced
      by $K_S^0$ and $\Lambda^0$ decays.
      These decays are
      reconstructed with the program {\sc Mammoth}~\cite{mammoth} for
      the data and detector simulation. At the generated level of
      the simulation, these particles have been
      treated as stable particles.

\item{\bf Secondary interactions} \\
     Another source of uncertainty stems from   particles
     produced in reinteractions
     of primary particles with the detector material.
     Positively charged pions and protons are produced in preference.
     All positively charged protons were omitted in the corresponding
     momentum range ($p \le 2.7$ \rm{GeV}/c) to study this effect.

\item{\bf Particle identification}  \\
     To take uncertainties of the particle identification into account,
     the results for different particle identification cuts
     (loose, standard, and tight) were compared.

\item{\bf Purity correction of the jets} \\
The flavour composition of the normal mixture sample has been varied by 
imposing cuts of different strength to the event sample. In this way three
samples were obtained, one with the flavour mix of Z decays, one which 
was depleted in b events and one depleted in b and c events.
The first and second sample were used to obtain pure ``duscb'' and ``dusc''
results and the third sample to obtain ``dusc'' and ``dus'' results using
the Monte Carlo. 
\end{enumerate}

Furthermore the results were compared to those obtained by using the {\sc
Cambridge} 
algorithm~\cite{cambridge} instead of the {\sc Durham} algorithm.
The change of the multiplicities then is typically 2\%; changes of the
ratios and double ratios are much smaller. 
Finally a systematic error of $\lesssim
2\%$ due to track reconstruction losses as determined from the overall
multiplicity measurements~\cite{overallmulti} is assumed. 
As both systematic errors discussed 
apply to particles in general, 
they are expected to cancel in the ratios and are therefore not included
in~\tref{t:partsys-y}.

All systematic errors discussed above apply to spectra of identified
particles. However, the statistical uncertainty here in is general much bigger
than the systematic error due to the binning of the data. 
Moreover many systematic uncertainties will cancel in the gluon to quark 
ratios which are the main subject of this paper. 
Systematic errors have therefore been neglected in the errors shown in the
particle distributions.

\BT
\renewcommand{\arraystretch}{1.3}
\begin{center}
\begin{tabular}{|c|c|c|c|c|}  \hline
Particle     & This Paper Y  & {\sc Delphi}~\cite{partlet} & {\sc OPAL} 
three-jet~\cite{Bagliesi:1998mp} & {\sc Argus}~\cite{l:argus:part} \\
\hline $\pi^\pm$    & \erg{0.997}{0.009}{0.013}  &  ---
             &\erg{1.016}{0.010}{0.010}   &  1 (def.) \\
$K^\pm$      & \erg{0.942}{0.019}{0.016}  &  \erg{0.930}{0.040}{0.020}
             &\erg{0.948}{0.017}{0.028}   &  $0.86\pm0.31$ \\
$p\bar{p}$   & \erg{1.205}{0.041}{0.025}  &  \erg{1.120}{0.110}{0.040}
             &\erg{1.100}{0.024}{0.027}   &  $1.58\pm0.10$ \\
\hline
\end{tabular}
\caption{$R'_X$ from measurements of different collaborations.}
\label{t:rprime_koll}
\end{center}
\ET

\BT[p]
\BC\small\renewcommand{\arraystretch}{0.93}
\BTB{|l|c|c|c|c|c|c|cccc|}\hline
\mc{11}{|c|}{Y events}\\\hline
          &           &           &      & Stat. & \mc{2}{|c|}{Summed Error}  & \mc{4}{|c|}{Syst. Error [\%] of} \\
\rb{Vari-}& \rb{Par-}& Quarks    & Value & Error&   with $K^0, \Lambda$  &   without $K^0, \Lambda$ &            & Sec. &  Part. & Purity-\\
\rb{able} & \rb{ticle} &           &      & [\%]  &   [\%]       &   [\%]       & \rb{$K^0, \Lambda$} & Int. &  Id.   & Corr.  \\\hline
               &         &duscb  & 5.85 & 0.62 & -& 1.21 & -& 1.01 & 0.60 & 0.29\\
               & $\pi$   &dusc   & 5.70 & 0.68 & -& 1.20 & -& 1.00 & 0.60 & 0.30\\
               &         &dus    & 5.67 & 0.71 & -& 1.19 & -& 1.00 & 0.60 & 0.19\\\cline{2-11}
               &         &duscb  & 0.74 & 1.63 & -& 1.77 & -& 0.95 & 1.49 & 0.14\\
               & $K$     &dusc   & 0.69 & 1.88 & -& 1.89 & -& 1.01 & 1.59 & 0.14\\
               &         &dus    & 0.64 & 2.04 & -& 1.84 & -& 0.94 & .57 & 0.16\\\cline{2-11}
\rb{$N_X^q$}   &         &duscb  & 0.33 & 3.01 & -& 1.13 & -& 0.90 & 0.30 & 0.60\\
               & $p$     &dusc   & 0.33 & 3.00 & -& 1.12 & -& 0.90 & 0.30 & 0.60\\
               &         &dus    & 0.34 & 3.21 & -& 1.09 & -& 0.87 & .29 & 0.58\\\cline{2-11}
               &         &duscb  & 7.08 & 0.44 & -& 1.01 & -& 1.00 & 0.06 & 0.06\\
               & $X^\pm$ &dusc   & 6.77 & 0.49 & -& 1.01 & -& 1.00 & 0.06 & 0.06\\
               &         &dus    & 6.65 & 0.51 & -& 1.01 & -& 1.01 & 0.06 & 0.05\\\hline\hline
               &         &duscb  & 7.07 & 0.45 & -& 1.09 & -& 1.00 & 0.40 & 0.17\\
               & $\pi$   &dusc   & 7.04 & 0.44 & -& 1.08 & -& 0.99 & 0.38 & 0.17\\
               &         &dus    & 7.05 & 0.45 & -& 1.08 & -& 1.01 & .38 & 0.11\\\cline{2-11}
               &         &duscb  & 0.84 & 1.19 & -& 1.75 & -& 0.95 & 1.43 & 0.36\\
               & $K$     &dusc   & 0.84 & 1.20 & -& 1.76 & -& 0.96 & 1.44 & 0.36\\
               &         &dus    & 0.84 & 1.20 & -& 1.74 & -& 0.96 & .44 & 0.24\\\cline{2-11}
\rb{$N_X^g$}   &         &duscb  & 0.49 & 1.86 & -& 1.68 & -& 1.03 & 1.03 & 0.82\\
               & $p$     &dusc   & 0.49 & 1.82 & -& 1.64 & -& 1.01 & 1.01 & 0.81\\
               &         &dus    & 0.49 & 1.84 & -& 1.57 & -& 1.02 & .02 & 0.61\\\cline{2-11}
               &         &duscb  & 8.57 & 0.30 & -& 1.01 & -& 1.00 & 0.01 & 0.08\\
               & $X^\pm$ &dusc   & 8.56 & 0.30 & -& 1.01 & -& 1.00 & 0.01 & 0.08\\
               &         &dus    & 8.56 & 0.30 & -& 1.01 & -& 1.00 & 0.01 & 0.04\\\hline\hline
               &         &duscb  & 1.21 & 0.75 & 1.68 & 1.14 & 1.24 & 0.99 & 0.25 & 0.50\\
               & $\pi$   &dusc   & 1.24 & 0.81 & 1.65 & 1.11 & 1.21 & 0.97 & 0.24 & 0.49\\
               &         &dus    & 1.24 & 0.80 & 1.58 & 1.02 & 1.21 & 0.97 & .24 & 0.24\\\cline{2-11}
               &         &duscb  & 1.14 & 2.02 & 1.65 & 1.63 & 0.26 & 0.96 & 1.31 & 0.09\\
               & $K$     &dusc   & 1.21 & 2.15 & 1.68 & 1.66 & 0.25 & 0.99 & 1.32 & 0.08\\
               &         &dus    & 1.31 & 2.37 & 1.68 & 1.64 & 0.31 & 0.99 & 1.30 & 0.08\\\cline{2-11}
\rb{$R_X$}     &         &duscb  & 1.46 & 3.42 & 2.13 & 2.11 & 0.27 & 1.03 & 0.96 & 1.58\\
               & $p$     &dusc   & 1.48 & 3.58 & 2.08 & 2.06 & 0.27 & 1.01 & 1.01 & 1.49\\
               &         &dus    & 1.43 & 3.57 & 1.99 & 1.97 & 0.28 & 0.98 & 0.98 & 1.40\\\cline{2-11}
               &         &duscb  & 1.21 & 0.50 & 1.19 & 0.99 & 0.66 & 0.99 & 0.08 & 0.00\\
               & $X^\pm$ &dusc   & 1.26 & 0.55 & 1.21 & 1.03 & 0.63 & 1.03 & 0.08 & 0.00\\
               &         &dus    & 1.29 & 0.62 & 1.19 & 1.01 & 0.62 & 1.01 & 0.08 & 0.00\\\hline\hline
               &         &duscb  & 1.00 & 0.90 & 1.27 & 1.12 & 0.60 & 1.00 & 0.30 & 0.40\\
               & $\pi$   &dusc   & 0.98 & 1.02 & 1.30 & 1.14 & 0.61 & 1.02 & 0.31 & 0.41\\
               &         &dus    & 0.97 & 1.04 & 1.26 & 1.19 & 0.62 & 1.04 & 0.31 & 0.21\\\cline{2-11}
               &         &duscb  & 0.94 & 2.02 & 1.73 & 1.51 & 0.85 & 0.96 & 1.17 & 0.11\\
\rb{$R'_X$}    & $K$     &dusc   & 0.96 & 2.20 & 1.89 & 1.64 & 0.94 & 1.05 & 1.26 & 0.10\\
               &         &dus    & 1.02 & 2.46 & 1.77 & 1.54 & 0.88 & 0.98 & 1.18 & 0.10\\\cline{2-11}
               &         &duscb  & 1.21 & 3.40 & 2.10 & 2.08 & 0.33 & 1.00 & 0.91 & 1.58\\
               & $p$     &dusc   & 1.17 & 3.67 & 2.10 & 2.07 & 0.34 & 1.02 & 0.94 & 1.54\\
               &         &dus    & 1.11 & 3.61 & 1.94 & 1.90 & 0.36 & 0.99 & 0.90 & 1.35\\\hline
\ETB

\EC
\caption{Systematic errors for $N_X^q$, $N_X^g$, $R_X$ and $R'_X$ in Y events.
         Here $X$ denotes the particle species $\pi, K, p$  or all charged
         particles. The summed error displays the quadratic sum of the
         individual errors.}
\label{t:partsys-y}
\ET

\subsection{Momentum Spectra}

\begin{figure}[p]
\breite 11.4cm
\breitemp 11.5cm
\begin{center}

\rotatebox{90}{
  \begin{minipage}[t]{\breitemp}
    \begin{center}
       \includegraphics*[angle=0,width=\breite]{\fileid{pa-p-spec-ai}}
       \caption{Momentum spectra of identified hadrons in \qgj
        a)-c) spectra of pions, kaons, and protons in
         quark jets; d)-f) corresponding spectra for
         gluon jets in events with Y topology.
         The predictions of the generator models
         \JS, \AR\ und \HW\ are drawn as lines.}
         \label{f:pa-p-spec}
    \end{center}
  \end{minipage}
\hspace{1cm}
  \begin{minipage}[t]{\breitemp}
    \begin{center}
      \includegraphics*[angle=0,width=\breite]{\fileid{pa-p-ratio-pop}}
      \caption{Ratios of the momentum spectra of identified hadrons
               in \gqj of Y events;
        a)-c) ratios of the spectra of pions, kaons, and protons in
        gluon jets to those in quark jets;
        d)-f) corresponding spectra normalized to the ratio
        gluon/quark for all charged particles.
        The predictions of the generator models
         \JS, \JS\ with default baryon production model
         and \HW\ are drawn as lines.}
        \label{f:pa-p-ratio}
    \end{center}
  \end{minipage}
}
\end{center}
\end{figure}

\fref{f:pa-p-spec} shows the momentum spectra of identified hadrons in
quark ($duscb$) and gluon jets for Y events.
The momentum spectra of kaons and protons differ significantly from those of
pions. Pions are produced mainly at low momentum, both in quark and gluon
jets. The likely explanation is that pions are often low
energy decay products of unstable particles.
The Monte Carlo generators {\sc Jetset}, {\sc Ariadne}, and {\sc Herwig}
describe the gross features of the measured {\sc Delphi} data.
The momentum distribution of kaons in gluon jets is best described by {\sc
Ariadne}. 
The {\sc Herwig} model shows a considerable weakness concerning
the description of kaon momentum spectrum in gluon jets. The multiplicity of
fast kaons is clearly underestimated. The momentum distribution of protons in
gluon jets is well modelled by the {\sc Jetset} and {\sc Ariadne} generators
but not by the {\sc Herwig} model.

\fref{f:pa-p-ratio} shows the ratios of the momentum spectra of
identified hadrons in \gqj.
This measurement is an improvement of
our previous publication~\cite{partlet}. More low energy particles are
produced in gluon jets than in quark jets for all kinds of particles.
At high particle momenta this structure is inverted.
\fref{f:pa-p-ratio}(d,e,f)  shows the corresponding ratios of the momentum
spectra of \fref{f:pa-p-ratio}(a,b,c) normalized to
the ratio of the momentum spectra of all charged particles in \gqj.

\fref{f:pa-p-ratio}(f) indicates that
the proton enhancement in gluon jets is bigger than
that for all charged particles.
The overestimate of the proton production ratio by {\sc Jetset} or {\sc
Ariadne} (not shown)
is presumably  due to an extra suppression of baryon production at the
end of the string (i.e. for quark jets, see also Section~\ref{sec_baryon}).
A much better description is obtained using the default baryon
production model without this extra suppression.

However,  no direct
conclusions concerning the ratios of the multiplicities can be
drawn from the normalized ratios as a function of momentum,
because the shapes of the momentum spectra of kaons and protons differ
significantly from those of pions which dominate the all charged particle
sample.

\subsection{Rapidity}

\begin{figure}[p]
\breite 11.4cm
\breitemp 11.5cm
\begin{center}

\rotatebox{90}{
  \begin{minipage}[t]{\breitemp}
    \begin{center}
       \includegraphics*[angle=0,width=\breite]{\fileid{pa-rap-ff}}
       \caption{Rapidity spectra of identified hadrons in \qgj compared to
                {\sc Jetset} in events with Y topology.}
       \label{f:pa-rap-FF}
    \end{center}
  \end{minipage}
\hspace{1cm}
  \begin{minipage}[t]{\breitemp}
    \begin{center}
      \includegraphics*[angle=0,width=\breite]{\fileid{pa-rap-ratio-pop}}
      \caption{Ratios of the rapidity spectra of identified hadrons in \gqj
               from Y events;
        a)-c) ratios of the spectra of pions, kaons, and protons in
        gluon jets to those in quark jets ;
        d)-f) corresponding spectra normalized to the ratio
        gluon/quark for all charged particles;
        The predictions of the generator models
        \JS, \JS\ with the default baryon production model,
        and \HW\ are drawn as lines.}
      \label{f:pa-rap-ratio}    \end{center}
  \end{minipage}
}
\end{center}
\end{figure}

\fref{f:pa-rap-FF}  shows the rapidity spectra of identified hadrons in \qgj
 and \fref{f:pa-rap-ratio}  shows the corresponding ratios.
For all particles there are in the plateau, i.e. at low $\eta$, ~1.6-2 times
more particles in gluon jets.
An excess of particles is expected due to the higher colour charge of the
gluon. 
At high $\eta$, i.e. in the range of the leading particle only few kaons and
protons are observed in gluon jets.

\subsection{$\xi$-Spectra}
\label{sec_baryon}

\begin{figure}[p]
\breite 11.4cm
\breitemp 11.5cm
\begin{center}

\rotatebox{90}{
  \begin{minipage}[t]{\breitemp}
    \begin{center}
       \includegraphics*[angle=0,width=\breite]{\fileid{pa-ksi-ff-ai-bb}}
       \caption{$\xi_p$ spectra of identified hadrons in \qgj compared to
                {\sc Jetset} in events with Y topology.}
       \label{f:pa-KSI-FF}
    \end{center}
  \end{minipage}
\hspace{1cm}
  \begin{minipage}[t]{\breitemp}
    \begin{center}
      \includegraphics*[angle=0,width=\breite]{\fileid{pa-qratios-upper-mc-ai}}
      \caption{Normalized ratio $R'_p = R_p / r_{ch}$
      for different compositions of the  quark event sample (Y events).
      }
      \label{f:pa-qratios}
    \end{center}
  \end{minipage}
}
\end{center}
\end{figure}

\fref{f:pa-KSI-FF}  shows the $\xi_p$ spectra of identified hadrons in
\qgj. The {\sc Jetset} and {\sc Ariadne} (not shown) models provide
a reasonable description over a wide range of the $\xi_p$ spectrum. 
The maximum height is different for quark and gluon jets
indicating different
particle rates. The point of intersection of the $\xi_p$ distributions of
\qgj\ for pions and kaons is approximately the same, $\xi^{(s)}_p \sim 1.73$.
For protons the crossing point between the quark and gluon distributions is
shifted to higher momentum at $\xi^{(s)}_p \sim 0.74$. Proton
production is enhanced in gluon jets, but preferentially at high momenta.
This can be seen more clearly in~\fref{f:pa-qratios} which
shows the normalized ratio
$R_p^{'}(\xi_p)$. It is observed that this ratio is unity within errors
at very small
$\xi_p$ (highest momenta) and close to unity  also at large $\xi_p$ (small
momenta). A strong deviation from unity is, however, visible in the
intermediate $\xi_p$ region ($0.7 \le \xi_p \le 1.85$).

\begin{figure}[t,b,h]
\centering\parbox{\bwid}{
\centering\includegraphics[width=\fwid]{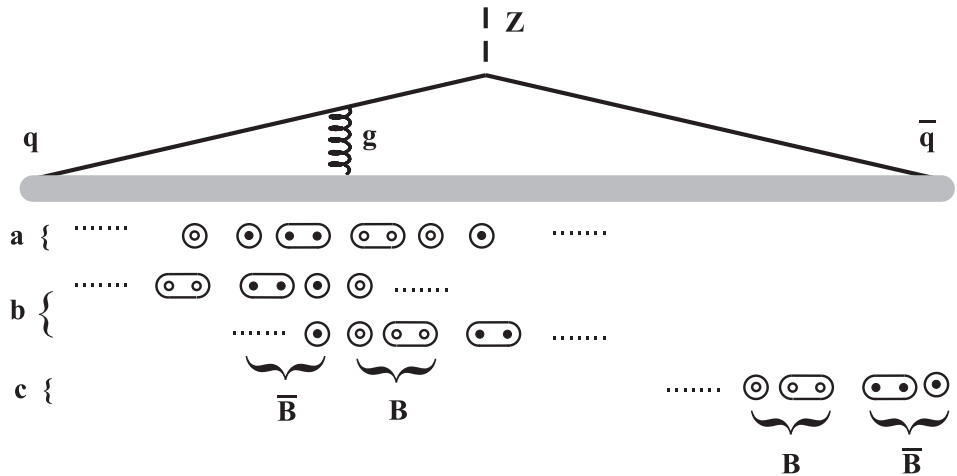}
\caption{\label{f:baryon}Different possibilities of
baryon production in strings. The single points
denote quarks and the double points diquarks. Open points stand for quarks and
filled points for anti-quarks. Line a\} illustrates a primary splitting into
diquark-anti-diquark in the center of the string. Line b\} shows the
possibilities for secondary diquark production. Line c\} shows a
diquark-anti-diquark splitting at a string end.}
}
\end{figure}

A surplus of baryon production in gluon jets and the observed
kinematical properties can be qualitatively understood if 
baryons are directly produced from coloured partons or
equivalently from a colour string.
In a parton shower colour conservation leads to the so-called
preconfinement property, that is a local compensation of colour charge
in space. Alternatively the produced colour charges can always be ordered to
form continuous chains or strings in space time.
These strings appear naturally in the Lund fragmentation model~\cite{lun1}
and in the progenitor model of Feynman and Field~\cite{fieldfeynman}.
A colour string ends at the primary quarks produced in the underlying
hard scattering but is spanned over the corresponding gluons.
Hadron production now can be assumed to proceed via a pair-creation of
a quark-anti-quark (or diquark-anti-diquark) pair and a corresponding
string break-up.
A single break-up in the vicinity of a primary quark will produce a
leading hadron, whereas close to the gluon in the centre of the string
at least two breaks are
needed before a hadron is formed.
To produce a baryon a production of a  diquark-anti-diquark pair is
compelling. Now it should be noted that 
in the centre of a string 
(i.e. in the vicinity of the gluon) more
possibilities exist which lead to baryon formation.
A primary diquark-anti-diquark break up as well as a secondary one
following a primary quark-anti-quark creation leads to baryon
production (see~\fref{f:baryon}).

\bwid 0.8\textwidth
\fwid 0.99\bwid

The latter process (marked b\} in~\fref{f:baryon}) may happen 
in any of the two remaining strings similarly to both
original endpoints of the string (see~\fref{f:baryon} c\}).
The first production mechanism (marked a\} in~\fref{f:baryon})
is missing at the string end.
This
leads to the excess of baryons in gluon compared to quark
jets. Here it is likely that two leading baryons are produced which take a
large fraction of the gluon energy. Thus it is expected that the
excess of baryon production centers at comparably large scaled
momentum which is indeed observed (see~\fref{f:pa-qratios}).
At small momentum, i.e. in the momentum range where baryons from the
inner part of the string between the jets are expected to contribute,
the relative portion of baryon
production in quark and gluon jets is approximately equal.

Although the above discussion centers around the string model it is based
on quite general topological properties and is a strong indication
that baryon production, and presumably also meson production, happens
directly from colour objects and an intermediate step of colour and
baryon number neutral objects is avoided.
In particular this is also indicated by the failure of the HERWIG
model to describe the surplus of proton production observed in the data.

In detail the above described
mechanism will be complicated by the abundant production
of resonant baryons~\cite{fred}-\cite{delta} or equivalently the
so-called popcorn-mechanism~\cite{lun1}.
Further support to this interpretation comes from the observed strong energy
dependence of the surplus of baryon production (compare the {\sc Argus}
measurement at $\sqrt{s}\simeq 10 \rm{GeV}$ 
to this result in~\tref{t:rprime_koll}).
The surplus of baryons in gluon jets is due to the leading baryons.
As energy and thus the multiplicity ratio in gluon to quark jets
increases~\cite{mulkhoze} this excess is less and less important in the double
ratio $R_p^{'}$. The Lund model qualitatively describes the decrease of
$R_p^{'}$ shown in~\tref{t:rprime_koll}.

\bwid 0.99\textwidth \fwid 0.8\bwid
\begin{figure}[htb]
\centering\parbox{\bwid}{
      \includegraphics*[angle=0,width=\fwid]{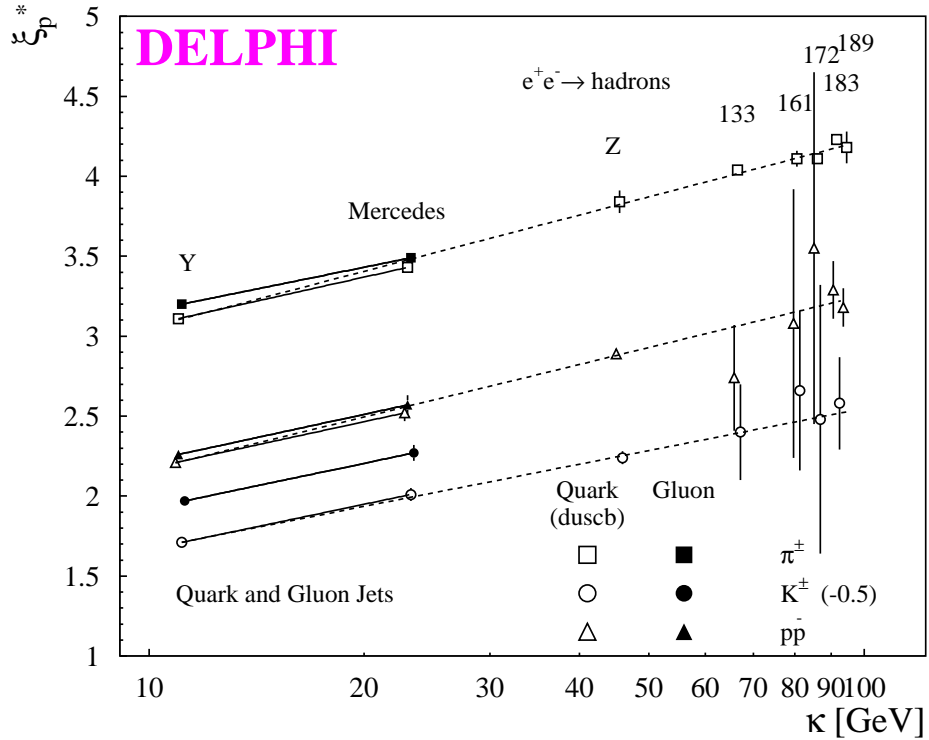}
      \caption{Fitted maxima of the $\xi_p$ spectra. 
               Here $\kappa = E_{jet}\sin\frac{\theta_1}{2}$
               and $\kappa=E_{beam}$ denote the underlying jet scale.
               Observe the offset of 0.5 units for the $\xi^*_p$ values
               for the $K^{\pm}$. The solid lines represent the fit to Y and
               Mercedes results, the dashed lines represent the fit
               including the $e^+e^-$ data.}
      \label{f:pa-ximax}
}
\end{figure}

\fref{f:pa-ximax} shows 
position of the maximum, $\xi_p^*$, of a simple Gaussian fitted to
the $\xi_p$ distributions
in dependence of the scale $\kappa = E_{jet} \cdot
\sin{\theta_{min}/2}$ with $\theta_{min}$ being the
angle with respect to the closest
jet, here $\theta_{min} = \theta_1$ (for a detailed discussion of jet scales
see~\cite{scaling}).
The fit results for $\xi_p^*$ are given in~\tref{t:ximax}. The maxima of the
$\xi_p$ distributions for protons and kaons for quark jets are shifted to
smaller values (i.e. higher momenta) compared to pions as has been observed
previously. It is clearly observed that the maximum is shifted to smaller
$\xi_p$ for $K$'s in quark compared to gluon jets. To a lesser extent this is
also observed for $\pi$'s and $p$'s. In MLLA/LPHD the maxima of the $\xi_p$
spectra for gluons and quarks are expected to be almost
identical~\cite{scaling,webber_book}.
A natural explanation for the observed difference especially for $K$'s is the
leading particle effect.
In~\cite{Abe:1999qh} it has been shown that in a scaled momentum range
corresponding to the
smallest $\xi$-bin in~\fref{f:pa-KSI-FF} leading charged kaons
(i.e. those containing a primary produced quark flavour) are about three
times more often produced than non-leading kaons.

\begin{table}
\begin{tabular}{|c|c@{$\,\pm\,$}c|c@{$\,\pm\,$}c|}   \hline
\mc{5}{|c|}{Y events} \\                   \hline
par- &  \mc{4}{|c|}{$\xi^*_p$} \\     \cline{2-5}
ticle&  \mc{2}{|c|}{quark} & \mc{2}{|c|}{gluon} \\ \hline
$\pi$   & 3.11 &0.01 &  3.20 & 0.00      \\
$K$     & 2.21 &0.02 &  2.47 & 0.01      \\
$p$     & 2.21 &0.02 &  2.26 & 0.01      \\
$X^\pm$ & 2.93 &0.01 &  3.05 & 0.00      \\
\hline
\end{tabular}
\hfill
\begin{tabular}{|c|c@{$\,\pm\,$}c|c@{$\,\pm\,$}c|}   \hline
\mc{5}{|c|}{Mercedes events} \\            \hline
par- & \mc{4}{|c|}{$\xi^*_p$} \\     \cline{2-5}
ticle& \mc{2}{|c|}{quark} & \mc{2}{|c|}{gluon} \\ \hline
$\pi$   & 3.43  & 0.03  & 3.49  & 0.02  \\
$K$     & 2.51  & 0.04  & 2.77  & 0.05  \\
$p$     & 2.52  & 0.05  & 2.57  & 0.06  \\
$X^\pm$ & 3.24  & 0.02  & 3.35  & 0.02  \\
\hline
\end{tabular}
\normalsize
\caption{Maxima of the  $\xi_p$ distributions. Errors are statistical only.}
\label{t:ximax}
\end{table}

For all particles the $\xi_p^*$ values are bigger for Mercedes than for Y
events, i.e. a scale evolution of the $\xi_p$ spectra is observed 
(see~\fref{f:pa-ximax} and~\tref{t:ximax}).
Assuming a general linear increase of
$\xi_p^*$ with the logarithm of the scale $\kappa$, i.e.
\[f(\xi_p^*) = a+b\ln(\kappa) \]
the quark jet measurements extrapolate reasonably well to the measurements
for
overall Z events~\cite{l:dis:jochen}, and
for high energy events~\cite{neufeld}.
The dashed  lines in~\fref{f:pa-ximax} indicate fits including the
Z and higher energy data.

It is further remarkable, that contrary to the predictions of the
LPHD model, the peak values of the $\xi_p$ distributions for kaons and protons
in quark jets are almost equal ($\xi^*_p$ values for the $K^{\pm}$
in~\fref{f:pa-ximax} are 
shifted by 0.5 units to avoid overlaps with the proton results.).
This observation contradicts the predictions of the
LPHD concept that the positions of the maxima of the $\xi_p$ distributions are
proportional to the logarithm of the mass of the corresponding particle.
It has  been shown already (see e.g.~\cite{l:dis:hahn}) that mesons and baryons
show a behaviour which differs from this simple expectation. This is a
consequence of heavy particle decays and of the partially different masses of
the decay particles in (predominantly baryon) decays.
This statement is qualitatively confirmed by this analysis.

\section{Summary and Conclusion}
\label{sec_sum}

Based on a sample of about 2.2 million hadronic \z decays collected
by the {\sc Delphi} detector at {\sc Lep},
the production of identified particles
in jets initiated by gluons or by quarks,
was analysed and compared.

As observed for inclusive charged particles,
the production spectrum of identified particles was found to be
softer in gluon jets compared to quark jets, with a higher total
multiplicity. The normalized multiplicity
ratio ($R'$) for protons in Y events was measured to be:
$$
 R'_{p} = \frac{R_{p}}{r_{ch}}= \frac{(N_p/N_{ch.})_g}{(N_p/N_{ch.})_q}
        =1.205 \pm 0.041_{stat.} \pm 0.025_{sys.} .
$$
$N_{p(ch)}$ denotes the number of protons (all charged particles).
{\sc Herwig} underestimates both the kaon and the proton
production in gluon jets.

This surplus of baryon production in gluon jets indicates that baryons are
produced directly from coloured partons or from strings
and that an intermediate state of neutral clusters (like in the {\sc Herwig}
cluster model) is avoided.
This interpretation is supported by the scaled energy dependence of the proton
excess and by the evolution of the proton excess with energy scale.

Furthermore the $\xi_p$ and $\eta$ distributions were measured and 
agreement with the {\sc Jetset} and {\sc Ariadne} models was found. 
{\sc Herwig} underestimates both the kaon and the proton
production in gluon jets.
The maxima of the $\xi$ distributions of quark jets,
$\xi_p^*$, extrapolate well with the scale $\kappa$ to those obtained from all
events at different centre-of-mass energies.
For kaons the maximum is shifted to smaller $\xi_p^*$ compared to gluon jets
presumably because of a leading particle effect.


\subsection*{Acknowledgements}
\hspace{14pt}

We thank T. Sj\"ostrand for useful and illuminating discussions,
especially for pointing out the possible explanation of the excess of baryon
production in gluon jets.\\
 We are greatly indebted to our technical
collaborators, to the members of the CERN-SL Division for the excellent
performance of the LEP collider, and to the funding agencies for their
support in building and operating the DELPHI detector.\\
We acknowledge in particular the support of \\
Austrian Federal Ministry of Science and Traffics, GZ 616.364/2-III/2a/98, \\
FNRS--FWO, Belgium,  \\
FINEP, CNPq, CAPES, FUJB and FAPERJ, Brazil, \\
Czech Ministry of Industry and Trade, GA CR 202/96/0450 and GA AVCR A1010521,\\
Danish Natural Research Council, \\
Commission of the European Communities (DG XII), \\
Direction des Sciences de la Mati$\grave{\mbox{\rm e}}$re, CEA, France, \\
Bundesministerium f$\ddot{\mbox{\rm u}}$r Bildung, Wissenschaft, Forschung
und Technologie, Germany,\\
General Secretariat for Research and Technology, Greece, \\
National Science Foundation (NWO) and Foundation for Research on Matter (FOM),
The Netherlands, \\
Norwegian Research Council,  \\
State Committee for Scientific Research, Poland, 2P03B06015, 2P03B1116 and
SPUB/P03/178/98, \\
JNICT--Junta Nacional de Investiga\c{c}\~{a}o Cient\'{\i}fica
e Tecnol$\acute{\mbox{\rm o}}$gica, Portugal, \\
Vedecka grantova agentura MS SR, Slovakia, Nr. 95/5195/134, \\
Ministry of Science and Technology of the Republic of Slovenia, \\
CICYT, Spain, AEN96--1661 and AEN96-1681,  \\
The Swedish Natural Science Research Council,      \\
Particle Physics and Astronomy Research Council, UK, \\
Department of Energy, USA, DE--FG02--94ER40817. \\

\clearpage


\end{document}